# Fractional Trigonometric Functions in Complex-valued Space:

## Applications of Complex Number to Local Fractional Calculus of Complex Function

Yang Xiao-Jun

*Department of Mathematics and Mechanics, China University of Mining and Technology, Xuzhou Campus, Xuzhou, Jiangsu, 221008, P. R.C*

dyangxiaojun@163.com

This paper presents the fractional trigonometric functions in complex-valued space and proposes a short outline of local fractional calculus of complex function in fractal spaces.

*Key words: Fractional trigonometric function, complex function, local fractional calculus, fractal space*

MSC2010：30E99, 28A80, 28A99

## Introduction 1

The trigonometric functions played an important role in both mathematics and engineering. Recently, the fractional trigonometric functions in real-valued space were discussed [1]. Recently, the fractional trigonometric functions in real-valued space were discussed in fractal space and their exponent was fractal dimension [2,3]. In similar manner the fractional trigonometric functions in complex-valued space were structured [2.3].

There are many definitions of local fractional calculus [2-11]. Hereby we write down Gao-Yang-Kang's local fractional derivatives [2-6]

$$f^{(\alpha)}(x_0) = \frac{d^\alpha f(x)}{dx^\alpha}\Big|_{x=x_0} = \lim_{x \to x_0} \frac{\Delta^\alpha(f(x) - f(x_0))}{(x - x_0)^\alpha}, \quad (1.1)$$

with $\Delta^\alpha(f(x) - f(x_0)) \cong \Gamma(1+\alpha)\Delta(f(x) - f(x_0))$.

and Gao-Yang-Kang's local fractional integrals [2-6]

$$_aI_b^{(\alpha)}f(x) = \frac{1}{\Gamma(1+\alpha)}\int_a^b f(t)(dt)^\alpha = \frac{1}{\Gamma(1+\alpha)}\lim_{\Delta t \to 0}\sum_{j=0}^{j=N} f(t_j)(\Delta t_j)^\alpha, \quad (1.2)$$

where $\Delta t_j = t_{j+1} - t_j$, $\Delta t = \max\{\Delta t_1, \Delta t_2, \Delta t_j, ...\}$, and $[t_j, t_{j+1}]$ for $j = 0, ..., N-1, t_0 = a, t_N = b$, is a partition of the interval $[a,b]$. Based on local fractional calculus, local fractional Fourier transforms [2] ,denoted by

$$f_\omega^{F,\alpha}(\omega) := \frac{1}{\Gamma(1+\alpha)}\int_{-\infty}^{+\infty} E_\alpha(-i^\alpha \omega^\alpha x^\alpha)f(x)(dx)^\alpha, 0 < \alpha \leq 1, \quad (1.3)$$

and local fractional Laplace transforms [3] , denoted by

$$f_s^{L,\alpha}(s) := \frac{1}{\Gamma(1+\alpha)} \int_0^{+\infty} E_\alpha(-s^\alpha x^\alpha) f(x)(dx)^\alpha, 0 < \alpha \leq 1, \qquad (1.4)$$

as new tools to deal with local fractional differential equations and local differential systems, were proposed. More recently, a new imaginary unit proposed in [2,3]. As a pursuit of the work we suggest fractional trigonometric functions in complex-valued space and their application to local fractional calculus of complex function.

## 2 The real-valued fractional trigonometric functions

In this section, we start with real-valued Mittag-Leffler function in fractal spaces. Here transforms method is proposed.

### 2.1 Mittag-Leffler function in fractal space

*Definition 1*

Let $E_\alpha : \mathbb{R} \to \mathbb{R}$, $x^\alpha \to E_\alpha(x^\alpha)$, denote a continuously function, which is so-called Mittag-Leffler function [2,3]

$$E_\alpha(x^\alpha) := \sum_{k=0}^{\infty} \frac{x^{\alpha k}}{\Gamma(1+\alpha k)}, 0 < \alpha \leq 1. \qquad (2.1)$$

**Remark 1**. The parameter $\alpha$ is fractal dimension. There always exists the relation

$$|E_\alpha(x^\alpha) - E_\alpha(y^\alpha)| < \zeta |x-y|^\alpha, \text{ for } x, y \in \mathbb{R},$$

where $\zeta$ is constant.

We have the following relations

$$E_\alpha(\lambda^\alpha x^\alpha) E_\alpha(\lambda^\alpha y^\alpha) = E_\alpha(\lambda^\alpha (x+y)^\alpha), \lambda \in C, \qquad (2.2)$$

and

$$E_\alpha(i^\alpha x^\alpha) E_\alpha(i^\alpha y^\alpha) = E_\alpha(i^\alpha (x+y)^\alpha), \qquad (2.3)$$

where the function $E_\alpha(i^\alpha x^\alpha)$ is periodic with the period $P_\alpha$ defined as the solution of the equation

$$E_\alpha(i^\alpha (P_\alpha)^\alpha) = 1, \qquad (2.4)$$

and

$$i^{2\alpha} = -1. \qquad (2.5)$$

As a direct result, we have [2]

$$E_\alpha(x^\alpha) E_\alpha(i^\alpha y^\alpha) = E_\alpha(x^\alpha + i^\alpha y^\alpha), \text{ for } 0 < \alpha \leq 1 \text{ and } x, y \in \mathbb{R}, \qquad (2.6)$$

Taking into account the relation (2.6) with $x = y = 0$, we arrive at the result

$$E_\alpha(0^\alpha) = 1.\qquad(2.7)$$

*Definition 2*

The fractional trigonometric function is denoted by

$$E_\alpha(i^\alpha x^\alpha) := \cos_\alpha x^\alpha + i^\alpha \sin_\alpha x^\alpha,\qquad(2.8)$$

with

$$\cos_\alpha x^\alpha := \sum_{k=0}^{\infty}(-1)^k \frac{x^{2\alpha k}}{\Gamma(1+2\alpha k)}\qquad(2.9)$$

and

$$\sin_\alpha x^\alpha := \sum_{k=0}^{\infty}(-1)^k \frac{x^{(2k+1)\alpha}}{\Gamma[1+\alpha(2k+1)]}.\qquad(2.10)$$

Successively, it follows from (2.9) and (2.10) that

$$\cos_\alpha 0^\alpha = 1 \qquad(2.11)$$

and

$$\sin_\alpha 0^\alpha = 0.\qquad(2.12)$$

**Remark 2.** Taking into account the fractal dimension $\alpha = 1$, the formulas (2.9) and (2.10) become respectively

$$\cos x = \sum_{k=0}^{\infty}(-1)^k \frac{x^{2k}}{\Gamma(1+2k)} \quad \text{and} \quad \sin x = \sum_{k=0}^{\infty}(-1)^k \frac{x^{2k+1}}{\Gamma[2(k+1)]}.$$

Hence, we have following result.

The function $E_\alpha(i^\alpha(P_\alpha)^\alpha)$ is periodic with the period $P_\alpha$ defined as the solution of the equation $E_\alpha(i^\alpha(P_\alpha)^\alpha) = 1$, then

$$P_\alpha = 2\pi.\qquad(2.13)$$

**2.2 Transforms method**

*Definition 3*

The circle of fractional order, which is defined by the equality

$$x^{2\alpha} + y^{2\alpha} = R^{2\alpha}, \quad x, y, R \in \mathbb{R},\ R > 0,\ 0 < \alpha \le 1.\qquad(2.14)$$

### Definition 4

The fractional-order circle region of order $\alpha$, $0 < \alpha \leq 1$, which is defined by the expression

$$x^{2\alpha} + y^{2\alpha} \leq R^{2\alpha}, \quad x, y, R \in \mathbb{R}, \ R > 0, \ 0 < \alpha \leq 1. \tag{2.15}$$

### Definition 5

The fractional-order equation of the roundness is defined by the equality

$$x^{2\alpha} + y^{2\alpha} + z^{2\alpha} = R^{2\alpha}, \quad x, y, z, R \in \mathbb{R}, \ R > 0, \ 0 < \alpha \leq 1. \tag{2.16}$$

### Definition 6

The fractional-order equation of the sphere is defined by the equality

$$x^{2\alpha} + y^{2\alpha} + z^{2\alpha} \leq R^{2\alpha}, \quad x, y, z, R \in \mathbb{R}, \ R > 0, \ 0 < \alpha \leq 1. \tag{2.17}$$

For (2.14) then there is a fractional-order trigonometric transform

$$\begin{cases} x^{\alpha} = R^{\alpha} \cos_{\alpha} \theta^{\alpha} \\ y^{\alpha} = R^{\alpha} \sin_{\alpha} \theta^{\alpha} \end{cases}, \tag{2.17}$$

where $0 < \theta < 2\pi$ and $R > 0$.

For (2.14) there is a fractional-order trigonometric transform

$$\begin{cases} u^{\alpha} = R^{\alpha} \sin_{\alpha} \eta^{\alpha} \cos_{\alpha} \theta^{\alpha} \\ v^{\alpha} = R^{\alpha} \sin_{\alpha} \eta^{\alpha} \sin_{\alpha} \theta^{\alpha} \\ w^{\alpha} = R^{\alpha} \cos_{\alpha} \eta^{\alpha} \end{cases}, \tag{2.19}$$

where $0 \leq \theta \leq 2\pi$ and $0 < \eta < \pi$.

## 3 The complex-valued fractional trigonometric functions

In this section, we start with complex-valued Mittag-Leffler function in fractal spaces.

### Definition 7

Let $E_{\alpha} : C \to C$, $z^{\alpha} \to E_{\alpha}(z^{\alpha})$, denote a continuously function, which is so-called the complex-valued Mittag-Leffler function

$$E_\alpha(z^\alpha) := \sum_{-\infty}^{+\infty} \frac{z^{k\alpha}}{\Gamma(1+\alpha k)}, \quad 0 < \alpha \leq 1, \tag{3.1}$$

**Remark 3.** The parameter $\alpha$ is fractal dimension. we always arrive at the relation

$$\left|E_\alpha(z_1^\alpha) - E_\alpha(z_2^\alpha)\right| < \zeta |z_1 - z_2|^\alpha, \text{ for } z_1, z_2 \in C,$$

where $\zeta$ is constant.

As a direct result, we have the following formulas:

$$E_\alpha(z_1^\alpha) E_\alpha(z_2^\alpha) = E_\alpha(z_1^\alpha + z_2^\alpha), z_1, z_2 \in C, \tag{3.2}$$

$$E_\alpha(z_1^\alpha) E_\alpha(z_2^\alpha) = E_\alpha((z_1 + z_2)^\alpha), z_1, z_2 \in C. \tag{3.3}$$

## Definition 8

A fractional-order complex number is given by

$$z^\alpha = x^\alpha + i^\alpha y^\alpha, \ z^\alpha \in C_1^\alpha, \ x, y \in \mathbb{R}, \ 0 < \alpha \leq 1, \tag{3.4}$$

its conjugate of complex number is denoted by

$$\overline{z^\alpha} = x^\alpha - i^\alpha y^\alpha, \ \overline{z^\alpha} \in C_1^\alpha, \ x, y \in \mathbb{R}, \ 0 < \alpha \leq 1, \tag{3.5}$$

and its fractional modulus is defined by the expression

$$\left|\overline{z^\alpha}\right| = |z^\alpha| = \sqrt{\overline{z^\alpha} \cdot z^\alpha} = \sqrt{x^{2\alpha} + y^{2\alpha}}. \tag{3.6}$$

It's easy to see that if $z^\alpha = x^\alpha + i^\alpha y^\alpha$ is purely real, that is, $\text{Re}(z^\alpha) = x^\alpha$. On the other hand, if $z^\alpha$ is purely imaginary, then $\text{Im}(z^\alpha) = y^\alpha$.

## Definition 9

The fractional trigonometric function is denoted by

$$E_\alpha(i^\alpha z^\alpha) := \cos_\alpha z^\alpha + i^\alpha \sin_\alpha z^\alpha, \tag{3.7}$$

with

$$\cos_\alpha z^\alpha := \sum_{k=0}^{\infty} (-1)^k \frac{z^{2\alpha k}}{\Gamma(1+2\alpha k)} \tag{3.8}$$

and

$$\sin_\alpha z^\alpha := \sum_{k=0}^{\infty} (-1)^k \frac{z^{(2k+1)\alpha}}{\Gamma[1+\alpha(2k+1)]}. \tag{3.9}$$

**Remark 4.** In special case of $\alpha = 1$ fractional-order complex number becomes

$$z = x + iy, \ z \in C, \ x, y \in \mathbb{R}, \tag{3.10}$$

its conjugate of complex number is denoted

$$\overline{z} = x - iy, \quad \overline{z^\alpha} \in C_1^\alpha, \; x, y \in \mathbb{R}, \tag{3.11}$$

yields the fractional modulus defined by the expression

$$|\overline{z}| = |z| = \sqrt{\overline{z} \cdot z} = \sqrt{x^2 + y^2}. \tag{3.12}$$

It follows the definition of classical complex number in special case of $\alpha = 1$.

**Theorem 1**

For a fractional-order complex number

$$z^\alpha = x^\alpha + i^\alpha y^\alpha, \; z^\alpha \in C_1^\alpha, \; x, y \in \mathbb{R}, \; 0 < \alpha \leq 1, \tag{3.13}$$

There exists an equivalent formula in the form of the trigonometric function, denoted by the expression

$$z^\alpha = x^\alpha + i^\alpha y^\alpha = \sqrt{x^{2\alpha} + y^{2\alpha}} \left( \cos_\alpha x^\alpha + i^\alpha \sin_\alpha x^\alpha \right). \tag{3.14}$$

Then

$$\cos_\alpha x^\alpha = \frac{x^\alpha}{\sqrt{x^{2\alpha} + y^{2\alpha}}} \tag{3.15}$$

and

$$\sin_\alpha x^\alpha = \frac{y^\alpha}{\sqrt{x^{2\alpha} + y^{2\alpha}}}. \tag{3.16}$$

*Proof.* Dividing by $\sqrt{x^{2\alpha} + y^{2\alpha}}$ in (3.14), we get

$$\frac{z^\alpha}{\sqrt{x^{2\alpha} + y^{2\alpha}}}$$
$$= \frac{x^\alpha}{\sqrt{x^{2\alpha} + y^{2\alpha}}} + i^\alpha \frac{y^\alpha}{\sqrt{x^{2\alpha} + y^{2\alpha}}} \tag{3.17}$$
$$= \cos_\alpha x^\alpha + i^\alpha \sin_\alpha x^\alpha.$$

Hence we deduce the result.

# 4 Application: Local fractional calculus of complex-variable function

In this section we give a short outline of local fractional calculus. It is a useful tool to deal with non-differentiable function in complex space.

## 4.1 Local fractional continuity of complex functions

Take into account the relation

$$\left|E_\alpha\left(z_1^\alpha\right)-E_\alpha\left(z_2^\alpha\right)\right|<\zeta\left|z_1-z_2\right|^\alpha, \qquad (4.1)$$

with any $z_1, z_2 \in z, z \in C$, $\zeta$ is constant,

which is called complex Hölder inequality of $E_\alpha\left(z^\alpha\right)$.

*Definition 11*

$$\left|f\left(z_1\right)-f\left(z_2\right)\right|<\zeta\left|z_1-z_2\right|^\alpha, \qquad (4.2)$$

with any $z_1, z_2 \in z, z \in C$, $\zeta$ is constant, $f(z)$ is complex Hölder function.

*Definition 12*

Given $z_0$ and $\left|z-z_0\right|^\alpha<\delta^\alpha$, then for any $z$ we have

$$\left|f(z)-f\left(z_0\right)\right|<\varepsilon^\alpha. \qquad (4.3)$$

Here $f(z)$ is called local fractional continuous at $z=z_0$, denoted by

$$\lim_{z\to z_0} f(z)=f\left(z_0\right). \qquad (4.4)$$

Setting for any $z\in C$ $f(z)$ is called local fractional continuous at $z$, $f(z)$ is called local fractional continuous on $C$, denoted by $f\in C_\alpha(C)$.

As a direct result, we have the following result:

Suppose that $\lim_{z\to z_0} f(z)=f\left(z_0\right)$ and $\lim_{z\to z_0} g(z)=g\left(z_0\right)$, then we have that

$$\lim_{z\to z_0}\left[f(z)\pm g(z)\right]=f\left(z_0\right)\pm g\left(z_0\right), \qquad (4.5)$$

$$\lim_{z\to z_0}\left[f(z)g(z)\right]=f\left(z_0\right)g\left(z_0\right), \qquad (4.6)$$

and

$$\lim_{z\to z_0}\left[f(z)/g(z)\right]=f\left(z_0\right)/g\left(z_0\right), \qquad (4.7)$$

the last only if $g\left(z_0\right)\neq 0$.

## 4.2 Local fractional derivatives of complex functions

Setting $F\in C_\alpha(C)$, the local fractional derivative of $F(z)$ at $z_0$ is

$$_{z_0}D_z^\alpha F(z)=:\lim_{z\to z_0}\frac{\Gamma(1+\alpha)\left[F(z)-F\left(z_0\right)\right]}{\left(z-z_0\right)^\alpha}, 0<\alpha\leq 1. \qquad (4.8)$$

If this limit exists, then the function $F(z)$ is said to be local fractional analytic at $z_0$, denoted by ${}_{z_0}D_z^\alpha F(z)$, $\left.\dfrac{d^\alpha}{dz^\alpha}F(z)\right|_{z=z_0}$ or $F^{(\alpha)}(z_0)$.

If this limit exists for all $z_0$ in a region $C_\alpha(C)$, then the function $f(z)$ is said to be local fractional analytic in a region $C_\alpha(C)$.

As a direct result for definition of local fractional derivatives, we have the following result:

Suppose that $f(z)$ and $g(z)$ are local fractional analytic functions, the following rules are valid:

$$\frac{d^\alpha(f(z)\pm g(z))}{dz^\alpha}=\frac{d^\alpha f(z)}{dz^\alpha}\pm\frac{d^\alpha g(z)}{dz^\alpha}; \tag{4.9}$$

$$\frac{d^\alpha(f(z)g(z))}{dz^\alpha}=g(z)\frac{d^\alpha f(z)}{dz^\alpha}+f(z)\frac{d^\alpha g(z)}{dz^\alpha}; \tag{4.10}$$

$$\frac{d^\alpha}{dz^\alpha}\left(\frac{f(z)}{g(z)}\right)=\frac{g(z)\dfrac{d^\alpha f(z)}{dz^\alpha}+f(z)\dfrac{d^\alpha g(z)}{dz^\alpha}}{g(z)^2}\quad\text{if }g(x)\neq 0; \tag{4.11}$$

$$\frac{d^\alpha(Cf(z))}{dz^\alpha}=C\frac{d^\alpha f(z)}{dz^\alpha}, \text{ where } C \text{ is a constant}; \tag{4.12}$$

If $y(z)=(f\circ u)(z)$ where $u(z)=g(z)$, then

$$\frac{d^\alpha y(z)}{dz^\alpha}=f^{(\alpha)}(g(z))\left(g^{(1)}(z)\right)^\alpha. \tag{4.13}$$

$$\frac{d^\alpha z^{k\alpha}}{dz^\alpha}=\frac{\Gamma(1+k\alpha)}{\Gamma(1+(k-1)\alpha)}z^{(k-1)\alpha}; \tag{4.14}$$

$$\frac{d^\alpha E_\alpha(z^\alpha)}{dz^\alpha}=E_\alpha(z^\alpha); \tag{4.15}$$

$$\frac{d^\alpha \sin_\alpha z^\alpha}{dz^\alpha}=\cos_\alpha z^\alpha; \tag{4.16}$$

$$\frac{d^\alpha \cos_\alpha z^\alpha}{dz^\alpha}=-\sin_\alpha z^\alpha. \tag{4.17}$$

### 4.3 Local fractional integrals of complex functions

Setting $f\in C_\alpha(C)$ and letting $f$ be defined, single-valued in $C$. The local fractional integral of $f(z)$ along the contour $C$ in $C$ from point $z_p$ to point $z_q$, is defined as

$$I_C^\alpha f(z) = \frac{1}{\Gamma(1+\alpha)} \lim_{|\Delta z| \to 0} \sum_{i=0}^{n-1} f(z_i)(\Delta z)^\alpha = \frac{1}{\Gamma(1+\alpha)} \int_C f(z)(dz)^\alpha, \qquad (4.18)$$

where for $i = 0, 1, \ldots, n$ $(\Delta z)^\alpha = z_i^\alpha - z_{i-1}^\alpha$, $z_0 = z_p$ and $z_n = z_q$.

For convenience, we assume that

$$_{z_0}I_{z_0}^{(\alpha)} f(z) = 0 \quad \text{if } z = z_0. \qquad (4.19)$$

Taking into account the definition of local fractional integrals, we have the following result:

Suppose that $f, g \in C_\alpha(C)$, the following rules are valid:

$$\frac{1}{\Gamma(1+\alpha)} \int_C (f(z) + g(z))(dz)^\alpha = \frac{1}{\Gamma(1+\alpha)} \int_C f(z)(dz)^\alpha + \frac{1}{\Gamma(1+\alpha)} \int_C g(z)(dz)^\alpha; \qquad (4.18)$$

$$\frac{1}{\Gamma(1+\alpha)} \int_C kf(z)(dz)^\alpha = \frac{k}{\Gamma(1+\alpha)} \int_C f(z)(dz)^\alpha, \text{ for a constant } k; \qquad (4.19)$$

$$\frac{1}{\Gamma(1+\alpha)} \int_C f(z)(dz)^\alpha = \frac{1}{\Gamma(1+\alpha)} \int_{C_1} f(z)(dz)^\alpha + \frac{1}{\Gamma(1+\alpha)} \int_{C_2} f(z)(dz)^\alpha, \qquad (4.20)$$

where $C = C_1 + C_2$;

### Theorem 2

If the contour $C$ has end points $z_p$ and $z_q$ with orientation $z_p$ to $z_q$, and if function $f(z)$ has the primitive $F(z)$ on $C$, then we have

$$\frac{1}{\Gamma(1+\alpha)} \int_C f(z)(dz)^\alpha = F(z_q) - F(z_p). \qquad (4.21)$$

*Proof.* The proof of the theorem is similar to that of real function and is omitted. For more detail for real function, see[4,6].

### Theorem 3

If $C$ is a simple closed contour, and if function $f(z)$ has a primitive on $C$ then

$$\frac{1}{\Gamma(1+\alpha)} \oint_C f(z)(dz)^\alpha = 0. \qquad (4.22)$$

*Proof.* The definition of a closed contour is that $z_q = z_p$. So

$$\frac{1}{\Gamma(1+\alpha)} \int_C f(z)(dz)^\alpha = F(z_q) - F(z_p) = 0. \qquad (4.23)$$

This proof of the theorem is completed.

*Corollary 4*

If the contours $C_1$ and $C_2$ have same end points and if $f(z)$ is local fractional analytic on $C_1$, $C_2$ and between them, then we have

$$\frac{1}{\Gamma(1+\alpha)}\int_{C_1} f(z)(dz)^\alpha = \frac{1}{\Gamma(1+\alpha)}\int_{C_2} f(z)(dz)^\alpha. \qquad (4.24)$$

*Proof.* If $C = C_1 - C_2$, then we have

$$\frac{1}{\Gamma(1+\alpha)}\int_{C} f(z)(dz)^\alpha = \frac{1}{\Gamma(1+\alpha)}\int_{C_1} f(z)(dz)^\alpha - \frac{1}{\Gamma(1+\alpha)}\int_{C_2} f(z)(dz)^\alpha = 0.$$

$$(4.25)$$

This proof of the corollary is completed.

**Corollary 5**

If the closed contours $C_1$, $C_2$ is such that $C_2$ lies inside $C_1$, and if $f(z)$ is local fractional analytic on $C_1$, $C_2$ and between them, then we have

$$\frac{1}{\Gamma(1+\alpha)}\int_{C_1} f(z)(dz)^\alpha = \frac{1}{\Gamma(1+\alpha)}\int_{C_2} f(z)(dz)^\alpha. \qquad (4.26)$$

*Proof.* Taking new same end points path and using *Corollary 4*, we deduce the result.